# A Rank-based Convex Hull method for Dense Data Sets

G.M. Megson[1], J.O. Cadenas[2]

Abstract

*A novel 2-D method for computing the convex hull of a sufficiently dense set of n integer points is introduced. The approach employs a ranking function that avoids sorting the points directly thus reducing the overall time complexity. The ranked points create a simple polygonal chain from which the Convex Hull can be found using a suitable O(n) method. The result is achieved by placing a bound on the density (or ratio) of points to m, where m is the maximum value of the ranking function required to represent the set of points yielding an O(n+m) method. A fast method is then developed based on the bit length, p, of the data set which reduces this time further. The required conditions are easily satisfied by image processing methods which determine the Hulls of polygonal regions where the densities are in the range of 3%. Our experiments on a range problem domains show that this is not atypical. Since the complexity of the method is related to the bit size p for current machines (p=32, p=64) the method is for all practical purposes O(n). A short proof is provided.*

1. Introduction

The convex hull of a finite set of points has been a long standing problem in computer science owing to its relevance in computer graphics, design automation, pattern recognition, and operations research [ 1-15]. The problem has a number of well-known approaches which are summarized below and the reader is referred to [3] for a comprehensive review. Interest in fast solutions to the convex hull problem has been renewed recently with the increased emphasis on visual security measures, the availability of sophisticated image processing on mobile devices, computer gaming [21], automatic collision detection techniques [20] and semantic web technologies to classify data.

In this paper we are concerned with the problem of generating the convex hull for a set of 2-D integer points (x,y) although the solution presented can easily be extended to fixed point representations by a suitable scaling and a reduction in density. The convex hull is the set of points, CH, which forms a perimeter containing all the other points. Alternatively, CH is a set of points which describes a polygon whose internal angles are each less than $180^0$. In other words all the interior points can be created from a linear combination of the set CH.

Chand and Kapur [23] presented one of the first solutions to the convex hull method using the 'gift-wrapping' approach which required $O(n^2)$ steps. Graham [1] described the first $O(n\log_2 n + Cn)$ method which essentially converted the 2-D points (x,y) to polar form (r, θ) and sorted on θ before scanning the list to determine the hull. Shamos [22] introduced several algorithms with $O(n\log_2 n)$ using Divide and Conquer approaches to partition the point set into subsets, compute the hull of the subsets and then merge the results to compute the final hull [19]. Again these methods produce $O(n\log_2 n)$ time complexity. Jarvis [29] reduced the time to O(nh) where h is the number of points on CH while output sensitive algorithms are known to compute the hull in O(nlog₂h) time, Chan[30]. The search for a fast algorithm exploits the fact that a known partial ordering of the data set allows further savings by reducing the need to consider points several times. Sklansky produced a O(n) method in 1972 [10] that was subsequently proved incorrect. McCallum and Vis produced a correct version in 1979 and a series of improvements Lee, 1983 [6], Preparata and Shamos, 1985 [9], among others culminated with the method due to Melkman, 1987 [8], here briefly reviewed in Section 2. Melkman's method is regarded as the most elegant because it requires minimal supporting apparatus and provides the hull in a single pass through the points.

---

[1] School of Electronics and Computer Science, University of Westminster, London, W1W 6XH.
[2] School of Systems Engineering, Reading University.

A characteristic and limiting factor in these approaches is the time spent forming partial or total orders on the list of points. The O(nlog₂n) methods require no restrictions on the initial point set because they produce a total order on the points before finding the hull while the O(n) methods assume that the point set forms a simple polygonal chain. The latter assumption controls the order of the points sufficiently to limit the scan options and avoids backtracking.

In this paper we introduce the idea of a ranking function that creates an ordering of the points to build a simple polygonal chain (see Section 3). The advantage of the ranking is that it produces the ordering faster than sorting and so produces the conditions for an O(n) method without a full sort, (see Section 4). As a result we produce a fast method that lifts the restriction imposed by previous methods and approaches O(n) (see Section 5) under conditions that are easily realised in practice as shown by the examples in Section 6.

## 2. Definitions and Preliminaries

A more formal definition of the convex hull is as follows (see [13, 15, 16] for introductions) :

Given a finite set S of size n =|S| with 2-D points $v_i=(x_i,y_i)$ the *convex hull* is the set CH of size h=|CH| of points $p_j=(x_j,y_j)$ that form a bounding polygon containing all the points in S.

The set CH is a special case of the affine hull of a set of points:

$$affine(S) = \left\{ \sum_{i=1}^{k} a_i\, p_i \,\middle|\, k > 0, p_i \in S, a_i \in R\,, \sum_{i=1}^{k} a_i = 1 \right\}$$

where the coefficients $a_i$ are confided to be nonnegative.

A *simple polygonal chain* is a graph with a set of vertices $w_i=(x_i,y_i)$ ordered such that $w_i w_{i+1}$ are edges linking the vertices together such that no edges intersect. It follows trivially that when a single edge is deleted from the cycle formed by the convex set CH it describes a simple polygonal chain when the vertices are scanned in clockwise (or counter clockwise) direction (see Figure 1).

A *bounding box* (or simply box), B, is a graph of four vertices ($b_1$, $b_2$, $b_3$, $b_4$) with the edges $b_1 b_2$, $b_2 b_3$, $b_3 b_4$, $b_4 b_1$, which contains CH and hence S in such a way that a subset of at least two of the points for n>2 in CH lie on the edges of B. These points are associated with the minimum and maximum x and y values drawn from the point set S. The length of the sides of B are defined by $m_1 = x_{max} - x_{min} + 1$ and $m_2 = y_{max} - y_{min} + 1$ with the number of points $m = m_1 * m_2$ or the area of B. Note that the definition does not demand that B encloses a rectangular region but this is the usual and easiest interpretation to work with and will be adopted for the rest of the paper.

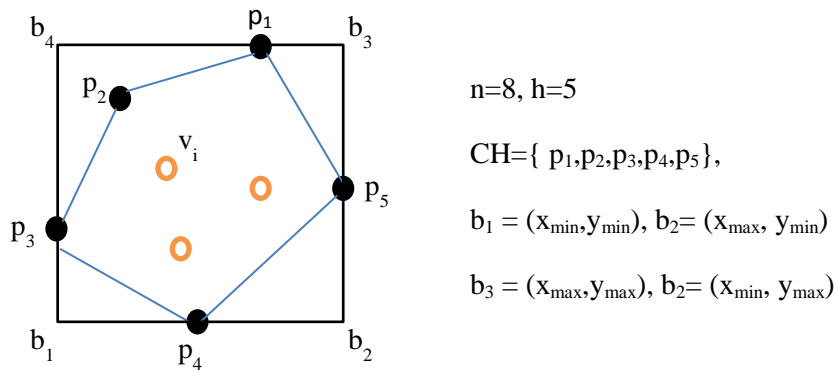

Figure 1: Convex Hull with Bounding Box

Given a set of three points it is useful to define a function *IsLeft*($v_0$, $v_1$, $v_2$) which returns True if the point $v_2$ lies to the left or right of the line formed by the edge $v_0v_1$. A simple way to perform the check is the formula:

$$g = (x_0 - x_1) * (y_2 - y_1) - (x_2 - x_1) * (y_0 - y_1)$$

expressed in programming notation and where $v_0=(x_0,y_0)$, $v_1=(x_1,y_1)$, $v_2=(x_2,y_2)$. The condition g>0 indicates that a point is IsLeft. If g=0 the three points are collinear.

A *ranking function* over the set of natural numbers, N, is defined as a mapping *f: N x N → N* such that $f(v_i)$ maps point $v_i$ to an integer in the range 1 ... m, for m>0. The value m is the maximum range of *f*() and we expect that m>n given the definition of the bounding box. The *inverse ranking function* is a mapping $f^{-1}$: *N → N x N* such that $f^{-1}(j)$ maps to a point $v_i$ with $f^{-1}(f(v_i)) = v_i$ and $f(f^{-1}(j)) = j$. It is not necessary in our method for the inverse ranking function to be in closed form and computable but it simplifies the operational details if these conditions are satisfied. A well-formed function f() maps each point in S to a unique value in the range 1...m and creates a simple polygonal chain. Additionally the computation time has to be comparable to IsLeft() for comparison purposes.

Next define the *density* of the point set S. Suppose that m is the number of points enclosed by the convex hull of the bounding box B then it is clear that each point of S is given a unique index by applying f(). Since m is also the area of the bounding box B we can define the ratio of points in the data set S to the points in the box B as D = n/m. Hence D represents how densely packed the box B is with respect to S. Since n = Dm where 0< D < 1 or m = kn where k>0 as n→m, k→1, density provides a way to measure if the size m is a simple linear scaling of n. The implication is that if the data is sufficiently dense the ranking method will remain O(n). Later we will relax this restriction.

### 3. A review of the Melkman Method

In 1987, Melkman provided an O(n) convex hull method, see [8] for full details. Here we present only a summary to capture the essence of the idea. Given the set of n points S arranged in a simple polygonal chain it is possible to traverse the point set and consider each point only once to determine if it is in the convex hull.

The constraint on the ordering of the vertices defines a window in which the next point of the hull must lie and so this simplifies the test for adding a point to the hull. The key to the algorithm is to define a Deque data structure which represents a partial chain of the convex hull. Suppose the deque is defined by *d* and the subscripts *t* and *b* define the top and bottom of the deque respectively. Elements can be introduced or removed by the corresponding operation of push/pop to the top or add/delete to the bottom. The top and bottom of the deque are identical at the end of each step representing that the hull is a simple chain that wraps around to form a cycle.

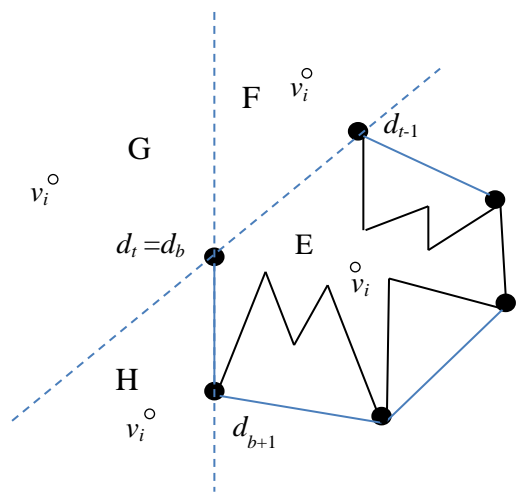

Figure 2 Exploration using Melkham's method

Figure 2, illustrates the basic idea. Scan the emerging hull anti-clockwise, then $d_b$, $d_{b+1}$, ... , $d_{t-1}$, $d_t$ is the state of the current set of points on the hull. Suppose that the next point to be considered is $v_i$. There are four possible regions to consider for the position of $v_i$, labelled E, F, G, H. The region E is confined to the area between the jagged line and the dotted lines $d_b d_{b+1}$ and $d_{t-1} d_t$ if this were not the case the set S would not be a simple chain. Hence the test

$$\text{IsLeft}(d_b, d_{b+1}, v_i) \text{ And } \text{IsLeft}(d_{t-1}, d_t, v_i)$$

determines if the point is in the hull or not. The point $v_i$ is not on the hull when the test is true since it lies inside the existing hull represented by the deque. If the point is in region F then IsLeft($d_b, d_{b+1}, v_i$) is True and IsLeft($d_{t-1}, d_t, v_i$) is False indicating that the top of the deque has to be replaced by $v_i$. This is done by popping the deque until the IsLeft() function becomes True and then pushing $v_i$. Likewise for region H, IsLeft($d_b, d_{b+1}, v_i$) = False and IsLeft($d_{t-1}, d_t, v_i$) = True indicating that the bottom of the deque should be replaced with $v_i$ which is achieved by deleting elements until the IsLeft() condition is satisfied and then adding $v_i$. Finally, if the point $v_i$ is in region G, then both tests fail and both the top and bottom elements of the deque need replacing which is achieved by working both ends of the deque until the IsLeft() conditions are satisfied.

The original paper provides a full analysis but since each point is considered only once and once placed on the Deque can only be deleted at most twice this implies that the method requires O(n) operations involving the IsLeft() function.

Hence the result of the current paper is to define a way to produce a simple chain in O(n). It then follows from the above argument that there exists a convex hull algorithm which is also O(n) and which places no restrictions on the point set making the technique generally applicable.

### 4. A simple Ranking Method

The idea of the ranking function is designed to first compute an ordering of the points and then shuffle the points to make them easily accessible to the Melkham method. The basis of this approach is to construct a pseudo-hash function which maps the points into a simple classification set in which the rank determines the position on the required polygonal chain. Since we are confined to 2-D this implies a way of traversing the bounding box B in a manner that describes a simple chain of points. There are numerous candidate functions and we only consider a simple ranking function to illustrate the method. Additional papers (in preparation) explore more sophisticated ranking.

Without loss of generalization, define the bounding box in a normalised form with the point $b_1$ = (1,1) and sides $m_1$ and $m_2$ so that $m = m_1 \cdot m_2$. Obvious choices for the ranking function are the simple indexing functions

$$f_1(i,j) = (i-1)*m_2 + j \quad \text{or} \quad f_2(i,j) = (j-1)*m_1 + i$$

used for mapping higher dimensional data into a one-dimensional form in a zig-zag pattern by scanning the verticals or horizontals of the box. The inverse functions exist and can be computed by using a look-up table or modular arithmetic with the form:

$$j = ((r-1) \% m_2) + 1 \quad \text{and} \quad i = (r-j) \text{ div } m_2 + 1$$

where $r = f(i,j)$, '%' is the remainder and div is the division operator returning an integer result.

The following lemmas establish that the ranking function (or zig-zag) produces a simple chain.

**Lemma 1:** Given $n < m$ integer points $v_i = (x_i, y_i)$ inside a bounding box of $m = m_1 * m_2$ points there exists a simple chain described by the mapping $f_1(i,j) = (i-1)*m_2 + j$.

**Proof: [by construction]**
Suppose each column of the bounding box contains $n_i$ points from S such that $n_1 + n_2 + \ldots + n_{m1}$ = n. Then the $n_i$ points on a given column can be joined in a simple chain. Let the first true point be at $(i, a_i)$ and the last at $(i, b_i)$ then the chain is formed by running from $(i, a_i)$ to $(i, b_i)$ connecting any points from S and bypassing points not in S. Adjacent columns i and i+1 are connected by an edge from $(i, b_i)$ to $(i+1, a_{i+1})$ for i= 1 ... $m_1$-1. This creates a simple chain covering all the point in S.

**Lemma 2:** Given $n < m$ integer points $v_i = (x_i, y_i)$ inside a bounding box of $m = m_1 * m_2$ points there exists a simple chain described by the mapping $f_2(i,j) = (j-1)*m_1 + i$.

**Proof: [by construction]**
Apply the reasoning in lemma 1 row-wise.

Thus we have established that the mapping function $f()$ preserves the polygonal chain property on the sparse set of points to be considered inside the box B compared to scanning the full set of points inside the box. This is a necessary condition of the ranking function.

For the next step define BLOOM as a one-dimensional Boolean array with elements indexed 1 ... m such that initially all the entries are 0 (False). Also define an identical sized array, INDIRECT, to hold the address (or index) of a point in S. The idea being that if BLOOM(j) is set INDIRECT(j) contains the index q of a point $v_q$ in S. Under these definitions we set

$$BLOOM(f(v_q))=1 \text{ and } INDIRECT(f(v_q)) = q$$

The names are indicative of the activity. The first array acts as a very simple Bloom filter [31] to represent the set of points S where there can be no false negatives or positives (by our above assumptions). It should be clear that when $f(v_q)=f_1(i,j)$ with $v_q=(i,j)$, BLOOM acts as an ordered set and running from 1...m defines a list of points indexed by INDIRECT. By Lemma 1 or 2 this forms a simple polygonal chain. The array INDIRECT also provides and inverse mapping so that $f^{-1}()\equiv INDIRECT$ with zero representing a null or undefined point in the set S. In principle a collocation or interpolation of INDIRECT can be used to build $f^{-1}()$. Also if the inverse function is known then we can dispense with INDIRECT completely by using $f^{-1}()$ as an indexing function. Neither of these possibilities will be expanded below for ease of exposition.

Next perform a shuffle operation that squeezes out the zero elements by moving all the valid index points to the front of INDIRECT. A simple routine for this is given below:

```
Procedure Shuffle(BLOOM, INDIRECT)
   Idx = 0
   For j = 1 to m
     If BLOOM(j) Then
       Idx = Idx + 1
       INDIRECT(Idx) = INDIRECT(j)
       If Idx = n then break loop
     End if
   Next j
End Proc
```

Notice that in this format the loop appears to take O(m) steps but will stop after n items have been located. Consequently the worst case of m steps occurs only when the last point is at the end of INDIRECT. The expected performance depends on the distribution of the points in the bounding box or on the efficiency of the ranking function and so the Shuffle time ranges between O(n) and O(m) but in any case is much faster than the IsLeft() step in the Melkham method.

An algorithm for computing the convex hull follows straightforwardly:

**Step 1:** find the maximum and minimum x and y in the points set S. (defines B and m)

**Step 2:** Translate the point set S to a set S' using $(x',y') = (x-x_{min}+1, y - y_{min} + 1)$

**Step 3:** For each point v in S' compute BLOOM(f(v)) and INDIRECT (f(v))

**Step 4:** Shuffle INDIRECT

**Step 5:** Apply the Melkham method with $v_i = v_{INDIRECT(j)}$ for j = 1 ... n from S

Step 1 requires a scan of n points so is O(n). Step 2 requires a translation for each point to simplify the computation of f() and is O(n). Step 3 requires n evaluations of the rank function f() so is O(n). Step 4 requires at most O(m) steps. And Step 5 is O(n) by discussion in the previous section. Since none of the operations in the above steps is more costly than IsLeft() we can conclude that the overall method is O(n + m), given that n ≤ m, the overall method is in fact O(m). However Step 4 can be accelerated significantly.

## 5. A Fast Shuffling method

We can do better by exploiting the fact that BLOOM can be used to rapidly skip gaps in INDIRECT to create a fast shuffle technique. To establish the principle first consider the problem of extracting only the 1-bits in a binary string [24]. In general a b-bit piece of data will require $2^b$ bits to represent all possibilities. Suppose that we wish to install the entries 8, 14, 4, 5, 2 into a set stored as a bit string, b=4 and we require a 16 bit number arranged as shown in Table 1.

| 15 | 14 | 13 | 12 | 11 | 10 | 9 | 8 | 7 | 6 | 5 | 4 | 3 | 2 | 1 | 0 |
|---|---|---|---|---|---|---|---|---|---|---|---|---|---|---|---|
| 0 | 1 | 0 | 0 | 0 | 0 | 0 | 1 | 0 | 0 | 1 | 1 | 0 | 1 | 0 | 0 |

Table 1 : a 16 bit number with positions 8, 14, 4, 5 and 2 set.

The string in Table 1 is the value 16692 in base 10 and we can extract the original data by applying the iteration procedure shown in Table 2 using the recurrence $x_j = x_{j-1}$ & $(x_{j-1} - 1)$, where $x_0$ = 16692, '&' is the bit-wise logical AND, and $s = \log(x_{j-1} - x_j)$.

| Iter. | x | $x_{j-1} - x_j$ | s |
|---|---|---|---|
| 0 | 16692 | | |
| 1 | 16688 | 4 | 2 |
| 2 | 16672 | 16 | 4 |
| 3 | 16640 | 32 | 5 |
| 4 | 16384 | 256 | 8 |
| 5 | 0 | 16384 | 14 |

Table 2: iteration process to extract index of 1-bits

**Theorem 1**: Given any positive integer $x$, where $n$ bits are set to one in a binary string of $k$ bits, $x = 0$ after $n$ iterations of the step defined by $x_j = (x_{j-1} -1)$ & $x_{j-1}$ the set bit positions are log $(x_{j-1} - x_j)$.

*Proof*: [by induction]

The definition of a binary integer representation for any positive integer x of $k$ bits is given by

$$x = b_{k-1}2^{k-1} + b_{k-2}2^{k-2} + \ldots + b_1 2^1 + b_0 2^0$$

Let the n bits that are set to one to be found at positions $p_r$ for $r= 1, 2, \ldots, n$, thus

$$x = 2^{p_n} + 2^{p_{n-1}} + \cdots + 2^{p_2} + 2^{p_1}$$

Basis: For $n = 1$, $x_0 = 2^{p_1}$ and $x_1 = 2^{p_1} \& (2^{p_1} - 1) = 0$ this follows by using two's complement arithmetic to compute $2^{p_1} - 1$ on a $k$ bit number. The 2's complement of (-1) has the form

$$(-1) = 2^{k-1} + 2^{k-2} + \ldots + 2^1 + 2^0$$

Hence

$$2^{p_1} + (-1) = 2^{p_1 - 1} + \cdots + 2^1 + 2^0$$

And it follows that $x_1=0$ since all the lower significant bits to $p_1$ sum to 1 and all bits after and including $p_1$ force a carry in the above expression. Now, $x_0 - x_1 = 2^{p_1}$ and $\log(x_0 - x_1) = p_1$

Next step: n=2 then $x_0 = 2^{p_2} + 2^{p_1}$ and $x_1 = x_0 \& (x_0 - 1)$ using the above definition of (-1)

$$x_0 + (-1) = 2^{p_2} + 2^{p_1} + (-1) = 2^{p_2} + 2^{p_1 - 1} + \cdots + 2^1 + 2^0$$

Since the bit a $p_1$ triggers a carry which then ripples towards bit $p_2$ which evaluates to 3 retaining bit $p_2$ and passing on the ripple. Hence it follows that $x_1$ contains only the non-zero bits in $x_0$ except for position $p_1$ which is set to zero. Thus bit $p_1$ has been eliminated with $x_1 = 2^{p_2}$ yielding the difference $x_0 - x_1 = 2^{p_1}$ and log $(x_0 - x_1) = p_1$.

Assumption: theorem holds for r=n-1 and test for r=n.

At step r=t by definition and the assumption

$$x_{t-1} = 2^{p_n} + 2^{p_{n-1}} + \cdots + 2^{p_t}$$

Hence

$$x_{t-1} + (-1) = 2^{p_n} + 2^{p_{n-1}} + \cdots + 2^{p_{t+1}} + 2^{p_t - 1} + \cdots + 2^0$$

and so

$$x_t = x_{t-1} \& (x_{t-1} + (-1)) = 2^{p_n} + 2^{p_{n-1}} + \cdots + 2^{p_{t+1}}$$

Yielding

$$x_{t-1} - x_t = 2^{p_t}$$

From which the result follows for r> 3. Hence true for all n.

**[end of proof]**

The above theorem allows a straightforward extension to the shuffle problem by observing that BLOOM is a bit string of length m and that extracting the 1-bits produces the indexing required for INDIRECT.  A revised version of Shuffle is given below:

```
Procedure Bit_Shuffle(BLOOM, INDIRECT)
   num = Convert(BLOOM)
   Idx = 0
   While num !=0 do
      x = num & (num-1)
      Idx = Idx + 1
      INDIRECT(Idx) = INDIRECT( log (num-x))
      num = x
    End if
   Wend
End Proc
```

The job of Convert is to simply generate a decimal number from the array BLOOM and from Theorem 1, it is clear that the while loop requires at most O(n) iterations. Unfortunately, the variable num requires m-bits and m≥n so although the performance is O(n) the data size depends on the size of the bounding box. In general the number of operations in Bits_Shuffle cannot be compared directly with IsLeft().

To address this problem, let p be the number of bits required to represent x or y of the data point (x,y) and hence the complexity of computations in IsLeft().The value p can be used to partition BLOOM into p-bit numbers. Let   r = ceil(m/p) then r represents the number of buckets required to quantise the ranking data in a way that assigns no more than p data points to each bucket. Each bucket then becomes a separate BLOOM vector which can be used to form a fast shuffle.

Load the data by defining BLOOM(0..r-1) as an array of p-bit numbers initially all set to zero (False). For each point $v_i$ in S compute u=Int(($f(v_i)$-1)/p)  and v = $2^{((f(vi)-1)\%p)}$ respectively with % the remainder operator. The value u is the bucket number where $f(v_i)$ index falls and v is the position within the p-bit number of the bucket that should be set (True) to represent the index $f(v_i)$.

$$BLOOM(u) = BLOOM(u) + v \text{ and } INDIRECT(f(v_i)) = i$$

For example, in Table 1, suppose bits are blocked together into groups of p=4, then r= m/p = 4. So BLOOM(0..3) is an array of 4-bit numbers initially all zero. Suppose we compute $f(v_i)$ = 5 then u = Int((5-1)/4) = 1 and v = $2^{((5-1)\%4)} = 2^0$ = 1 hence

$$BLOOM(1) = BLOOM(1)+ 1 \text{ and } INDIRECT(5) = i$$

With BLOOM(1) = $0001_2$. Likewise for $f(v_j)$ = 8, u=Int((8-1)/4) = 1, v=$2^{(7\%4)} = 2^3$ = 8

$$BLOOM(1) = BLOOM(1) + 8 \text{ and } INDIRECT(8) = j$$

Where $BLOOM(1) = 1001_2$ showing that $BLOOM(1)$ has ranked two elements. For $f(v_q) = 2$ we have $u = Int((2-1)/4) = 0$, and $v = 2^{((2-1)\%4)} = 2^1 = 2$ hence

$$BLOOM(0) = BLOOM(0) + 2 \text{ and } INDIRECT(2) = q$$

With $BLOOM(0) = 2 = 0010_2$ showing that the element with rank 2 has been stored correctly. With the ranking stored in this way applying Theorem 1 on each BLOOM element produces the following:

```
Procedure Fast_Shuffle(BLOOM, INDIRECT)
   Idx = 0
   For j = 0 to r-1
     num = BLOOM(j)

     While num !=0 do
       x = num & (num-1)

       Idx = Idx + 1
       INDIRECT(Idx) = INDIRECT(p*j + log (num-x))
       num = x
     End if
   Wend
  Next j

End proc
```

Observe that Fast_Shuffle uses data of size p-bits rather than the m-bits used in Bit_Shuffle and so is comparable to the IsLeft() operations in the Melkham method. The outer loop executes $r = Int(m/p)$ times and the while loop executes once for each bit set in BLOOM(j). By the construction of BLOOM we see that the n points in S are distributed across the buckets and since there are no duplicates the bits can only be set once. Hence if $t_j$ is the number of bits set in bucket j

$$n = t_0 + \cdots \cdots + t_{r-1}$$

**Theorem 2:** Fast_Shuffle has complexity $O(n)$ when $n \geq r$ and $O(n + C(m/p - n))$ when $n < r$.

**Proof [by exhaustion of cases]**

Case 1 ($n \geq r$) : The term $C(m/p-n)$ is bounded by n and effectively disappears. By the pigeon hole principle we know each BLOOM vector has at most $ceil(n/p)$ slots. Assuming a uniform distribution of points such that each BLOOM vector contains some real data the result follows straightforwardly since $m/p < n$. Next consider a case when some BLOOM vectors are zero. For example say $m = 64$ and $p = 16$ so $r = m/p = 4$, for $n = 37$ ($n > r$) and suppose BLOOM = [16, 16, 5, 0] we still have to check the last BLOOM entry, so we have $O(n + m/p - ceil(n/p))$. Since $p>1$, $ceil(n/p) < n$ likewise, $m/p < n$ and so is still $O(n)$ overall. Since there are less than n zero

BLOOM values this argument can be applied inductively on increasing the number of empty buckets.

Case 2 (n<r): The m/p-n additional operations are the tests for when BLOOM(j) =0. Since such a test can be computed in constant time of at most p-bit checks (this is O(1)), the coefficient C<1 represents the scaling compared to performing an IsLeft() operation. By definition we know that IsLeft() requires 2 multiplications and 5 subtract operations and then a test similar to the BLOOM test. In the Melkham method there at least two tests of IsLeft() to check the membership of a point on the hull and possibly more if the Deque has to be re-organised. A standard multiplication requires $O(p^2)$ and addition and subtraction requires $O(p)$ bit operations. Consequently IsLeft() has a time related to $2p^2 + 5p + 1$ and $C \approx 1/[2(2p^2 + 5p + 1)]$ counting in terms of Isleft() applications. Hence, for a point set of size n, bounding box size m, with computations performed on p-bit numbers we can write the time for Fast_Shuffle as:

$$O(n + \frac{1}{2(2p^2+5p+1)}(\frac{m}{p} - n))$$

And so the case holds when for some small constant, k,

$$\frac{1}{2(2p^2 + 5p + 1)}\left(\frac{m}{p} - n\right) < kn$$

**[end of proof]**

Below we use this fact to produce an upper bound on checking the zero BLOOM entries using p and the density of the point set and show that this is satisfied under quite ordinary circumstances.

The apparently more costly operation found in Fast_Shuffle, the log function, is easily computed by hardware support for bit operations found in most major processor manufacturers. For instance, from table 2, s = 4 is calculated directly by applying the so called intrinsic operation of count trailing zeros (ctz, see [25]) to the bit pattern "10000" ($16_{10}$) at iteration 2. This operation is accepted directly by the compiler, so the explicit call to a log() function disappears. We have validated the Fast_Shuffle procedure using the ctz instruction and it is much faster than an explicit call to log() and comparable to the basic operations in IsLeft().

**6. Complexity argument**

The above discussion establishes that the ranking approach is O(n) when n = m/p since the second term vanishes. Likewise when n > m/p the shuffle technique cannot execute more than n steps and again the second term is removed. This leaves only the case n <m/p to consider. Recalling the definition of density we see that:

$$\frac{n}{m} < \frac{1}{p} = D$$

This implies that the algorithm starts to drift away from linear performance when density of points drops below the reciprocal of the bit-length used in the computation. For current computing equipment bit lengths are of the order 32 or 64 bits. This suggests that more than 1 in 32 or 1 in 64 of the points inside the bounding box need to be members of the point set for the rank method to perform in linear time.

The latter is a conservative result since imposing the inequality from theorem 2:

$$\frac{1}{2(2p^2 + 5p + 1)}\left(\frac{m}{p} - n\right) < n$$

Also provides linearity. In this case a simple re-organisation yields

$$\frac{1}{2p(2p^2 + 5p + 1) + 1} < \frac{n}{m} = D$$

Again for p=32, D>0.0000071= $7.1 \times 10^{-6}$ and for p=64, D>00000092=$9.2 \times 10^{-7}$ which is a density of more than one in a hundred thousand or one in a million points. So for practical purposes we conclude that the presented algorithm is effectively linear.

| Field of Application | Size of m (typical) | Density range |
|---|---|---|
| Visual Hulls (skull images) [ref] | 3.2 – 3.6 x$10^6$ | 32% - 79% |
| Home range of wildlife animals [ref] | Any to 1.6x$10^8$ | 3% - 56% |
| Mass estimation of mammals [ref] | 4.1 - 8.9x$10^6$ | 24% - 54% |

Table 3. Typical densities of convex hulls

Typical densities evaluated from five different areas that exploit convex hulls are shown in Table 3. Figure 3 shows the running time of steps 1 through 4 proposed here to generate the convex hull for different ranking values of common image sizes of $m_{small}$ =640*480 and $m_{big}$ = 2048*1536. The number of points n has been varied from 256 points up to a density of 85% for $m_{small}$ and 10% for $m_{big}$. The Block size is p = 32. Each experiment shows a linear trend.

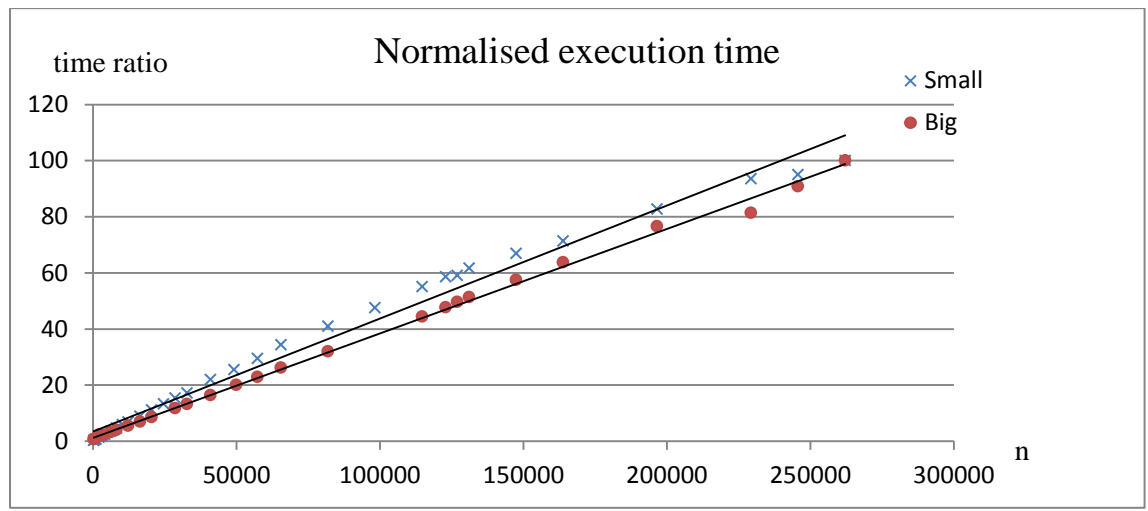

Figure 3. Normalized execution time for different n sizes on a small and big m.

The ranking time percentage to the total execution time also has a linear trend, although it appears to flatten to around 15% when n = 8192 for $m_{small}$ and n = 65536 for $n_{big}$. These two points are consistent to n > r (=m/p) as predicted. It can also be seen that these values correspond roughly to D = 1/p of around 3%.

We analysed the effect of the blocking size p. For this we maintained m fixed at m = 640*480 while we varied n from densities below 1% to around 42% for three natural p sizes of 16, 32 and 64. We used the counting trailing zero intrinsic function (ctz) in the experiments and found that p = 64 is always a better option resulting in improvements ranging from around 5% to 40% compared to p = 32 although the improvement decreases for larger densities. The test architecture was a 64-bit machine.

## 7. Discussion

There are also other ways to ensure that the algorithm performs in a linear time. A simple approach is to choose a ranking function that minimises the difference m-n. Clearly if the ranking function is perfect in the sense that m=n then D=1 and the above conditions are immediately satisfied. Another approach is to relax the restriction on the bounding box, B. A rectangular shape is easy to scan and so index but in principle all that is required is for the method to produce a shape that encloses the points S. In principle any affine set would suffice provided we apply a generating function for the interior points in which the complexity of the function is limited to a simple multiple of IsLeft().

The choice of ranking function can also help in minimising the amount of backtracking (or Deque manipulation) of points in the Melkman method by scanning in a manner that maximises the size of the emerging hull. The authors have explored a number of functions with these properties and will report the outcomes in follow-up papers. We mention two here for completeness. The Cantor counting technique scans a 2-D area by diagonals which opens up the hull to the widest extend as quickly as possible provided the points are mapped into a single quadrant by a suitable transformation like step 2 of the algorithm. An integer spiral also enumerates the 2-D plane but cannot be used as a ranking function without modification because the distribution of n points along the spiral of length m does not produce a simple chain of points. That is, there is a function and an inverse mapping but the necessary condition (see lemma 1 and 2) does not apply for spirals in general.

By finding a box of m points that encloses all n 2-D points, where m > n, this method is faster than any $O(nlog_2n)$ convex hull method when ratios $n/m \geq 1/log_2n$. In fact, this paper argues that the method approaches O(n) due to typical densities n/m of practical applications of convex hulls. The method is outlined as five steps, using the Melkman O(n) convex hull method as the last step. The first steps creates a simple polygonal chain required for Melkman to proceed without the need of an explicit sorting of the points.

## 8. Conclusion

In this paper we have introduced the idea of a ranking function to provide a total order of points without sorting by comparison thus circumventing the $O(nlog_2n)$ limit for finding the convex hull of a 2-D set on an unordered set of point of size n. The range of the ranking function is expressed as the area of a bounding box (m) and the density D=n/m used to construct and analyse a fast ordering (shuffle) method to prepare the points for consumption by the well-known O(n) method due to Melkham.

By using the complexity of the basic bit operations associated with the Melkham method we were able to show that the density properties required for O(n) performance are satisfied in quite ordinary circumstances. Application of the technique to large image databases provided empirical evidence that the method does indeed produce linear performance.

Finally, in this text, we explored only a simple ranking function. We are aware that many other such functions exist. We will publish analysis of several more sophisticated functions shortly. However it remains a challenge to construct a general function that is efficient in the sense that it minimises the metric m-n>0 and provides the tightest fitting bounding box and hence highest density.

**References**

[1]   R.L. Graham, An efficient algorithm for determining the convex hull of a finite planar set, Info. Proc. Lett. 1, 132-133 (1972).
[2]   M. Andrew. Another efficient algorithm for convex hulls in two dimensions. Inform. Process. Lett., 9(5):216-219, 1979.
[3]   Greg Aloupis, A History of Linear-time Convex Hull Algorithms for Simple Polygons, McGill Univ website (2000)
[4]   Bykat, "Convex hull of a finite set of points in two dimensions", Info. Proc. Letters 7, 296-298 (1978)
[5]   Ronald Graham & F. Yao, "Finding the convex hull of a simple polygon", J. Algorithms 4(4), 324-33 (1983)
[6]   D.T. Lee, "On finding the convex hull of a simple polygon", Int'l J. Comp. & Info. Sci. 12(2), 87-98 (1983)
[7]   D. McCallum and D. Davis, "A linear algorithm for finding the convex hull of a simple polygon", Info. Proc. Letters 9, 201-206 (1979)
[8]   Melkman, "On-line construction of the convex hull of a simple polygon", Info. Proc. Letters 25, 11-12 (1987)
[9]   Franco Preparata & Michael Shamos, Computational Geometry: An Introduction, Section 4.1.4 "Convex hull of a simple polygon" (1985)
[10]  Sklansky J., "Measuring Concavity on a Rectangular Mosaic", IEEE Transactions on Computing 21, p-1355 (1972).
[11]  Andrew, A. M. (1979), "Another efficient algorithm for convex hulls in two dimensions", *Information Processing Letters* **9** (5): 216–219, doi:10.1016/0020-0190(79)90072-3 .
[12]  Brown, K. Q. (1979), "Voronoi diagrams from convex hulls", *Information Processing Letters* **9** (5): 223–228, doi:10.1016/0020-0190(79)90074-7 .
[13]  de Berg, M.; van Kreveld, M.; Overmars, Mark; Schwarzkopf, O. (2000), *Computational Geometry: Algorithms and Applications* , Springer, pp. 2–8, http://books.google.com/books?hl=sv&lr=&id=tkyG8W2163YC&oi=fnd&pg=PA2 .
[14]  Chazelle, Bernard (1993), "An optimal convex hull algorithm in any fixed dimension" , *Discrete and Computational Geometry* **10** (1): 377–409, doi:10.1007/BF02573985 , http://www.cs.princeton.edu/~chazelle/pubs/ConvexHullAlgorithm.pdf .
[15]  Grünbaum, Branko (2003), *Convex Polytopes*, Graduate Texts in Mathematics (2nd ed.), Springer, ISBN 9780387004242.
[16]  Knuth, Donald E. (1992), *Axioms and hulls* , Lecture Notes in Computer Science, **606**, Heidelberg: Springer-Verlag, p. ix+109, doi:10.1007/3-540-55611-7 , ISBN 3-540-55611-7, MR 1226891 , http://www-cs-faculty.stanford.edu/~uno/aah.html .

[17] Krein, M.; Šmulian, V. (1940), "On regularly convex sets in the space conjugate to a Banach space", *Annals of Mathematics*, 2nd ser. **41**: 556–583, doi:10.2307/1968735 , JSTOR 1968735, MR 2009

[18] Schneider, Rolf (1993), *Convex bodies: The Brunn–Minkowski theory*, Encyclopedia of Mathematics and its Applications, **44**, Cambridge: Cambridge University Press, ISBN 0-521-35220-7, MR 1216521

[19] F.P. Preparata and S. Hong, "Convex hulls of finite sets of points in two and three dimensions",*Comm. ACM*, Vol. 20, 1977, pp. 87-93.

[20] T. Lozano-Perez, "An algorithm for planning collision-free paths among polyhedral obstacles",*Comm. ACM*, Vol. 22, 1979, pp. 560-570.

[21] R.O. Duda and P.E. Hart, *Pattern Classification and Scene Analysis*, Wiley, New York,1973.

[22] M.I. Shamos, "Computational geometry", Ph.D. thesis, Yale University, 1978.

[23] Chand, D.R. Kapur, S.S, An algorithm for convex polytopes, J. ACM 17, 7, (1970), 78-86

[24] B. W. Kernighan and D. Ritchie, *The C Programming Language*, (2[nd] Ed) Prentice Hall, New Jersey, 1988.

[25] http://gcc.gnu.org/onlinedocs/gccint/Misc.html Accessed on: 12-09-2012

[26] J. S. Franco and E. Boyer, "Exact Polyhedral Visual Hulls", British Machine Vision Conference, Vol. I, pp. 329-338, September 2003.

[27] http://gcc.gnu.org/onlinedocs/gccint/Misc.html and http://locoh.cnr.berkeley.edu/ Accessed on: 29-09-2012.

[28] W. I. Sellers et. al, "Minimum convex hull mass estimations of complete mounted skeletons", Biol. Lett. 2012;doi:10.1098/rsbl.2012.0263:1-4.

[29] Jarvis, R. A. (1973). "On the identification of the convex hull of a finite set of points in the plane". *Information Processing Letters* **2**: 18–21. doi:10.1016/0020-0190(73)90020-3

[30] Timothy M. Chan. "Optimal output-sensitive convex hull algorithms in two and three dimensions", *Discrete and Computational Geometry*, Vol. 16, pp.361–368. 1996

[31] B. H. Bloom. "Space /Time Trade-offs in Hash Coding with Allowable Errors". *Communication of the ACM*, vol. 13, pp. 422–426, July 1970.